\documentclass{PoS}
\usepackage{amsmath}
\usepackage{cite}
\usepackage[utf8]{inputenc}

\newcommand{\bq}{\begin{eqnarray}}
\newcommand{\eq}{\end{eqnarray}}
\newcommand{\eps}{\varepsilon}

\title{From elliptic curves to Feynman integrals}

\ShortTitle{From elliptic curves to Feynman integrals}

\author{Luise Adams\\
        PRISMA Cluster of Excellence, Johannes Gutenberg-Universit\"at Mainz\\
        E-mail: \email{ladams01@uni-mainz.de}}

\author{Ekta Chaubey\\
        PRISMA Cluster of Excellence, Johannes Gutenberg-Universit\"at Mainz\\
        E-mail: \email{eekta@uni-mainz.de}}

\author{\speaker{Stefan Weinzierl}\\
        PRISMA Cluster of Excellence, Johannes Gutenberg-Universit\"at Mainz\\
        E-mail: \email{weinzierl@uni-mainz.de}}

\abstract{
In this talk we discuss Feynman integrals which are related to elliptic curves.
We show with the help of an explicit example that in the set of master integrals more than one elliptic curve
may occur.
The technique of maximal cuts is a useful tool to identify the elliptic curves.
By a suitable transformation of the master integrals the system of differential equations 
for our example can be brought into a form linear in $\varepsilon$,
where the $\varepsilon^0$-term is strictly lower-triangular.
This system is easily solved in terms of iterated integrals.
}

\FullConference{Loops and Legs in Quantum Field Theory (LL2018)\\
		29 April 2018 - 04 May 2018\\
		St. Goar, Germany}

\begin{document}

\section{Introduction}

By an ``elliptic'' Feynman integral we understand a Feynman integral, which (i) cannot be expressed in terms of multiple polylogarithms
and (ii) is related to one or more elliptic curves (but not to any more complicated geometric object).
These Feynman integrals start at two-loops and play an important role for precision calculations with massive particles for LHC phenomenology.
They are a current topic of 
research 
\cite{MullerStach:2011ru,Adams:2013nia,Bloch:2013tra,Remiddi:2013joa,Adams:2014vja,Bloch:2014qca,Sogaard:2014jla,Adams:2015gva,Adams:2015ydq,Bloch:2016izu,Remiddi:2016gno,Adams:2016xah,Bonciani:2016qxi,Passarino:2016zcd,vonManteuffel:2017hms,Primo:2017ipr,Adams:2017ejb,Bogner:2017vim,Ablinger:2017bjx,Remiddi:2017har,Lee:2017qql,Bourjaily:2017bsb,Hidding:2017jkk,Broedel:2017kkb,Broedel:2017siw,Adams:2018yfj,Broedel:2018iwv,Groote:2018rpb,Adams:2018bsn,Adams:2018kez,Lee:2018ojn,Broedel:2018rwm,Adams:2018ulb}.
Elliptic Feynman integrals, which are related to a single elliptic curve are for example the massive sunrise integral or the kite integral.
In this talk we focus on a more complicated example, the planar double box integral for top-pair production with a closed top-loop \cite{Adams:2018bsn,Adams:2018kez}.
This integral enters the NNLO contribution for the process $p p \rightarrow t \bar{t}$.
In the existing NNLO calculation for the process $pp \rightarrow t \bar{t}$ this integral has been treated 
numerically \cite{Czakon:2013goa,Baernreuther:2013caa,Czakon:2008zk,Czakon:2007wk,Czakon:2005rk}, since it has not been known analytically.
Our lack of an analytic answer impedes further progress on the analytical side and motivates our study of this integral.
In this talk we will discuss how this integral can be treated analytically.

\begin{figure}
\begin{center}
\includegraphics[scale=1.0]{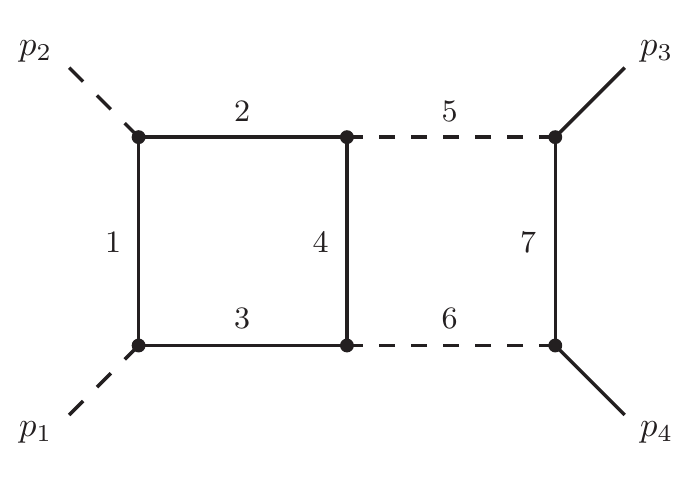}
\end{center}
\caption{
The planar double box integral. Solid lines correspond to massive propagators, dashed lines to massless propagators.
}
\label{fig1}
\end{figure}

The planar double box integral is shown in fig.~\ref{fig1}.
We take all momenta to be outgoing. The on-shell conditions for the external particles are
\bq
 p_1^2 \; = \; p_2^2 \; = \; 0,
 & &
 p_3^2 \; = \; p_4^2 \; = \; m^2.
\eq
The Mandelstam variables are defined by $s=(p_1+p_2)^2$ and $t=(p_2+p_3)^2$.
We consider the integral
\bq
 I_{\nu_1 \nu_2 \nu_3 \nu_4 \nu_5 \nu_6 \nu_7}\left( D, \frac{s}{m^2}, \frac{t}{m^2} \right)
 =
 e^{2 \gamma_E \eps}
 \left(m^2\right)^{\sum\limits_{j=1}^7 \nu_j - D}
 \int \frac{d^Dk_1}{i \pi^{\frac{D}{2}}} \frac{d^Dk_2}{i \pi^{\frac{D}{2}}}
 \prod\limits_{j=1}^7 \frac{1}{ P_j^{\nu_j} },
\eq
where $D=4-2\eps$ denotes the space-time dimension, $\gamma_E$ Euler's constant and the convention for the propagators is $P_j = -k_j^2 + m_j^2$.
We are interested in the Laurent expansion in $\eps$ of $I_{1111111}$ (and all its sub-topologies).
We show that these can be computed 
systematically to all orders in $\eps$ in terms of iterated integrals.

Let us briefly review iterated integrals \cite{Chen}.
For differential 1-forms $\omega_1$, ..., $\omega_k$ on a manifold $M$
and a path $\gamma : [0,1] \rightarrow M$ let us write 
for the pull-back of $\omega_j$ to the interval $[0,1]$
\bq
 f_j\left(\lambda\right) d\lambda & = & \gamma^\ast \omega_j.
\eq
The iterated integral is defined by
\bq
 I_{\gamma}\left(\omega_1,...,\omega_k;\lambda\right)
 & = &
 \int\limits_0^{\lambda} d\lambda_1 f_1\left(\lambda_1\right)
 \int\limits_0^{\lambda_1} d\lambda_2 f_2\left(\lambda_2\right)
 ...
 \int\limits_0^{\lambda_{k-1}} d\lambda_k f_k\left(\lambda_k\right).
\eq
Multiple polylogarithms are a special case of iterated integrals. Here, all integration kernels are given by
\bq
 \gamma^\ast \omega_j & = & \frac{d\lambda}{\lambda-c_j}.
\eq
The solution of the Feynman integrals $I_{\nu_1 \nu_2 \nu_3 \nu_4 \nu_5 \nu_6 \nu_7}$ reduces to multiple
polylogarithms for $t=m^2$.

A second special case are iterated integrals of modular forms.
If $f(\tau)$ a modular form, we will simply write $f$ instead of $2 \pi i f d\tau$ in the arguments of iterated
integrals.
The solution of the Feynman integrals $I_{\nu_1 \nu_2 \nu_3 \nu_4 \nu_5 \nu_6 \nu_7}$ reduces to
iterated integrals of modular forms of $\Gamma_1(6)$ for $s=\infty$.

For the analytic treatment of the planar double box integral we proceed in three steps:
In the first step we derive the differential equation in a pre-canonical basis. This step is in principle standard,
however there are some subtleties.
In the second step we transform the differential equation into a form linear in $\eps$,
where the $\eps^0$-term is strictly lower-triangular.
This is the essential step.
In the third step we solve the differential equation order by order in $\eps$.
Due to the second step this is easy.

\section{The system of differential equations}

The technique of differential equations \cite{Kotikov:1990kg,Kotikov:1991pm,Remiddi:1997ny,Gehrmann:1999as,Argeri:2007up,MullerStach:2012mp,Henn:2013pwa,Henn:2014qga,Tancredi:2015pta,Ablinger:2015tua,Adams:2017tga,Bosma:2017hrk}
is a powerful tool for Feynman integral calculations.
We may derive the differential equations as follows: 
Let us denote the first and second graph polynomials by ${\mathcal U}$ and ${\mathcal F}$, respectively.
Let us further denote by ${\mathcal F}_s$ and ${\mathcal F}_t$ the terms of the second graph polynomial
proportional to $(-s)$ and $(-t)$, respectively.
We introduce the dimensional shift operators ${\bf D}^\pm$, which shift the space-time dimension by two units
and propagator raising operator ${\bf j}^+$.
With these definitions, the dimensional shift relation reads
\bq
\label{dimensional_shift}
 {\bf D}^- I_{\nu_1 \nu_2 \nu_3 \nu_4 \nu_5 \nu_6 \nu_7}
 & = &
 {\mathcal U}\left( \nu_1 {\bf 1}^+, ..., \nu_7 {\bf 7}^+ \right)
 I_{\nu_1 \nu_2 \nu_3 \nu_4 \nu_5 \nu_6 \nu_7}.
\eq
For the derivatives we have
\bq
 \frac{d}{ds} I_{\nu_1 \nu_2 \nu_3 \nu_4 \nu_5 \nu_6 \nu_7}
 & = &
 {\bf D}^+
 {\mathcal F}_s\left( \nu_1 {\bf 1}^+, ..., \nu_7 {\bf 7}^+ \right)
 I_{\nu_1 \nu_2 \nu_3 \nu_4 \nu_5 \nu_6 \nu_7},
 \nonumber \\
 \frac{d}{dt} I_{\nu_1 \nu_2 \nu_3 \nu_4 \nu_5 \nu_6 \nu_7}
 & = &
 {\bf D}^+
 {\mathcal F}_t\left( \nu_1 {\bf 1}^+, ..., \nu_7 {\bf 7}^+ \right)
 I_{\nu_1 \nu_2 \nu_3 \nu_4 \nu_5 \nu_6 \nu_7}.
\eq
We may apply these two equations to the elements of a pre-canonical basis $\vec{I}$ 
with higher powers of the propagators, but no numerators.
Supplementing the equations with integration-by-parts identities 
and the inverse relation of eq.~(\ref{dimensional_shift}) gives us the sought after differential equation:
\bq
 d \vec{I} & = & A \vec{I}.
\eq
The connection one-form $A$ has to satisfy the integrability condition
\bq
 d A & = & A \wedge A.
\eq
The derivation sketched above may not be the most efficient method to derive the system of differential equations, but it is a very robust method,
requiring as input only integration-by-parts identities which reduce integrals to master integrals.
These can be obtained with standard programs like 
{\tt Reduze} \cite{vonManteuffel:2012np},
{\tt Kira} \cite{Maierhoefer:2017hyi} or
{\tt Fire} \cite{Smirnov:2014hma}.
We computed the integral reductions with all three programs in early 2018, using the current versions at this time.
Taking trivial symmetry relations into account and not using any advanced options, all programs gave 45 master integrals.
However, we observed that
the reductions for the three most complicated topologies seem to disagree
and that the results of two of the three programs seem to fail the integrability check.
Investigating this problem we discovered that there is an additional relation, which reduces
\begin{figure}
\begin{center}
\includegraphics[scale=1.0]{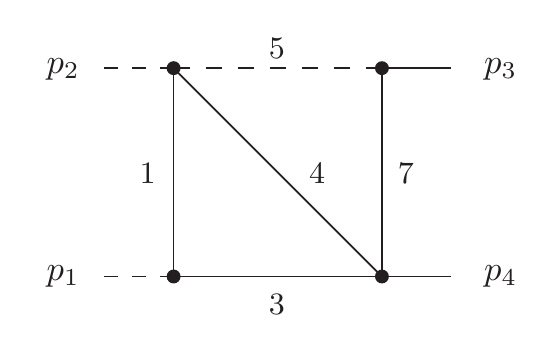}
\end{center}
\caption{
The topology, where the additional relation reduces the number of master integrals.
}
\label{fig2}
\end{figure}
the number of master integrals in the topology shown in fig.~\ref{fig2} from $5$ to $4$ and in turn the total
number of master integrals from $45$ to $44$.
This extra relation comes from a higher sector and reads
\bq
 0 & = &
 3\, \left( D-4 \right)
  \left( {m}^{2}-s-t \right)  \left( {m}^{2}-t \right) I_{1011101}
 +
 2\, 
  \left( {m}^{2}-s-t \right)  \left( {m}^{2}-t \right) {m}^{2} I_{2011101}
 \nonumber \\
 & &
 +
 2\, 
  \left( {m}^{2}-s-t \right)  \left( {m}^{2}-t \right) {m}^{2} I_{1021101}
 +
  \left( 2\,{m}^{4}-3\,{m}^{2}s-2\,{m}^{2}t+st \right)  \left( {m}^{2}-t \right) I_{1012101}
 \nonumber \\
 & &
 +
 4\, 
  {s}^{2} {m}^{2} I_{1011201}
 +
 \mbox{sub-topologies}.
\eq
Taking this additional relation into account, all integral reductions from the three programs are consistent and correct.
We would like to mention that {\tt Reduze} is able to find the relation 
and can be forced to use this relation with the command
\verb|distribute_external|\footnote{We thank L. Tancredi and A. von Manteuffel.}.
We also would like to mention that the current version $1.1$ of {\tt Kira} 
gives $44$ master integrals\footnote{We thank P. Maierhoefer and J. Usovitsch.}.

\section{Basis transformation}

In a second step we seek a transformation
\bq
 \vec{J} & = & U \vec{I},
\eq
such that the transformed differential equation is linear in $\eps$
\bq
\label{linear_in_eps}
 d \vec{J}
 & = &
 \left( A^{(0)} + \eps A^{(1)} \right) \vec{J},
\eq
where $A^{(0)}$ and $A^{(1)}$ are $\eps$-independent,
$A^{(0)}$ is strictly lower triangular
and $A^{(1)}$ is as usual block triangular.
The differential equation in eq.~(\ref{linear_in_eps})
can be brought into an $\eps$-form, if one introduces primitives for the entries of $A^{(0)}$.

Let us now introduce dimensionless variables $x$ and $y$ through
\bq
 \frac{s}{m^2} = - \frac{\left(1+x^2\right)^2}{x\left(1-x^2\right)},
 & &
 \frac{t}{m^2} = y.
\eq
The definition of $x$ simultaneously rationalises the square roots
\bq
 \sqrt{-s(4m^2-s)} & \mbox{and} & \sqrt{-s(-4m^2-s)}.
\eq
The second square root 
\begin{figure}
\begin{center}
\includegraphics[scale=1.0]{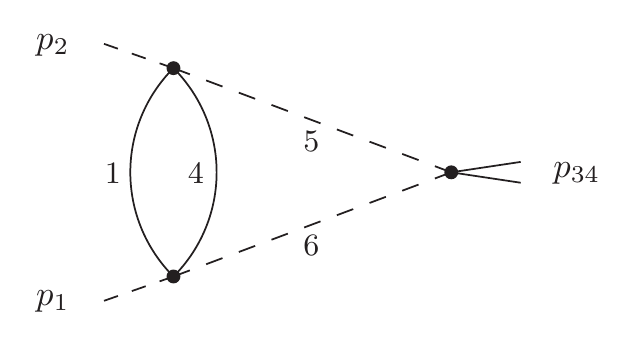}
\end{center}
\caption{
A Feynman graph giving rise to the square root $\sqrt{-s(-4m^2-s)}$.
}
\label{fig3}
\end{figure}
first enters through the Feynman integral shown in fig.~\ref{fig3}.
All sub-topologies, which depend only on $s/m^2$ (and all integrals in the limit $y=1$)
can be expressed as multiple polylogarithms with letters given
by 
\begin{alignat}{3}
 &
 \omega_0
 & \; = \; &
 \frac{ds}{s} 
 & \; = \; &
 \frac{2 dx}{x-i}
 +
 \frac{2 dx}{x+i}
 - \frac{dx}{x-1} - \frac{dx}{x+1} - \frac{dx}{x},
 \nonumber \\
 &
 \omega_4
 & \; = \; &
 \frac{ds}{s-4m^2}
 & \; = \; &
 \frac{2 dx}{x-\left(1+\sqrt{2}\right)}
 +
 \frac{2 dx}{x-\left(1-\sqrt{2}\right)}
 - \frac{dx}{x-1} - \frac{dx}{x+1} - \frac{dx}{x},
 \nonumber \\
 &
 \omega_{-4}
 & \; = \; &
 \frac{ds}{s+4m^2}
 & \; = \; &
 \frac{2 dx}{x-\left(-1+\sqrt{2}\right)}
 +
 \frac{2 dx}{x-\left(-1-\sqrt{2}\right)}
 - \frac{dx}{x-1} - \frac{dx}{x+1} - \frac{dx}{x},
 \nonumber \\
 &
 \omega_{0,4}
 & \; = \; &
 \frac{ds}{\sqrt{-s\left(4m^2-s\right)}}
 & \; = \; &
 \frac{dx}{x-1} - \frac{dx}{x+1} + \frac{dx}{x},
 \nonumber \\
 &
 \omega_{-4,0}
 & \; = \; &
 \frac{ds}{\sqrt{-s\left(-4m^2-s\right)}}
 & \; = \; &
 -\frac{dx}{x-1} + \frac{dx}{x+1} + \frac{dx}{x}.
 \nonumber
\end{alignat}
As an example consider the Feynman integral shown in fig.~\ref{fig12}.
\begin{figure}
\begin{center}
\includegraphics[scale=1.0]{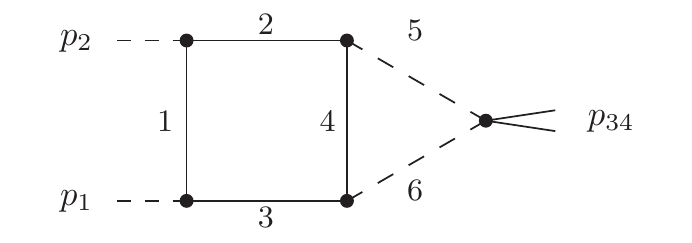}
\end{center}
\caption{
A Feynman integral evaluating to multiple polylogarithms.
}
\label{fig12}
\end{figure}
This integral yields
\bq
 J_{36}
 & = & 
 \eps^4 
 {\frac { \left( {x}^{2}+1 \right) ^{3} \left( {x}^{2}-2\,x-1 \right) }{ \left( x-1 \right) ^{2} \left( x+1 \right) ^{2}{x}^{2}}}
 I_{1111110}
 \\
 & = &
 \left[
 2\,I\left( \omega_{0,4}, \omega_{0,4}, \omega_{0}, \omega_{0,4}; x \right)
+2\,I\left( \omega_{0,4}, \omega_{0,4}, \omega_{0,4}, \omega_{0}; x \right)
-7\,I\left( \omega_{0,4}, \omega_{0}, \omega_{0,4}, \omega_{0,4}; x \right)
 \right.
 \nonumber \\
 & &
 \left.
+4\,I\left( \omega_{0,4}, \omega_{-4,0}, \omega_{-4,0}, \omega_{0}; x \right)
-4\,\zeta_2\,I\left( \omega_{0,4}, \omega_{-4,0}; x \right)
-10\,\zeta_3\,I\left( \omega_{0,4}; x \right)
-{\frac {39}{2}}\,\zeta_4
 \right] \eps^4
 + O\left(\eps^5\right). 
 \nonumber
\eq
There are a few integrals, which only depend on the variable $t$. 
\begin{figure}
\begin{center}
\includegraphics[scale=0.8]{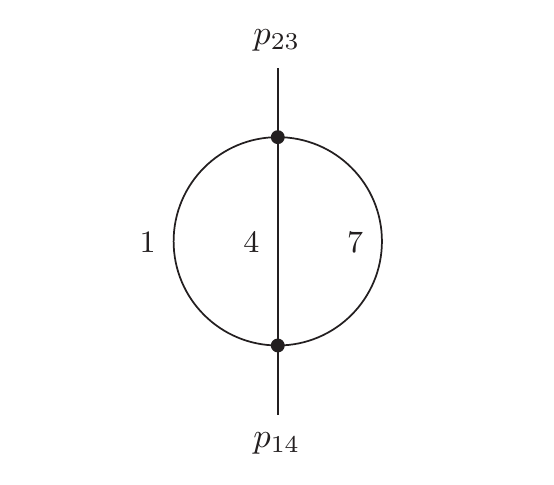}
\includegraphics[scale=0.8]{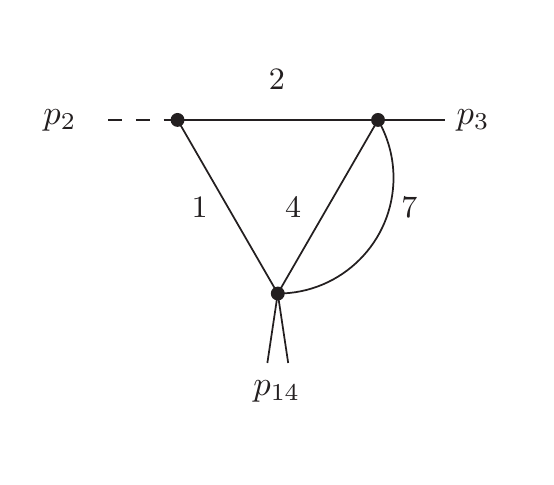}
\includegraphics[scale=0.8]{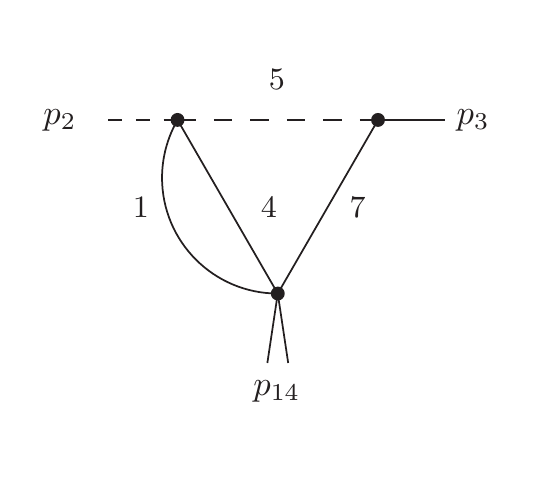}
\end{center}
\caption{
Feynman integrals depending only on $t$, but not on $s$.
}
\label{fig456}
\end{figure}
These are shown in fig.~\ref{fig456}.
These integrals (and all integrals in the limit $s=\infty$) can be expressed in terms 
of iterated integrals of modular forms of $\Gamma_1(6)$. The set of modular forms is given by
\bq
 \left\{ 1, g_{2,0}, g_{2,1}, g_{2,9}, g_{3,1}, p_{3,0}, g_{4,0}, g_{4,1}, g_{4,9}, p_{4,0}, p_{4,1} \right\},
\eq
where
\bq
 g_{n,r} = - \frac{1}{2} \frac{y\left(y-1\right)\left(y-9\right)}{y-r} \left( \frac{\psi^{(a)}_1}{\pi} \right)^n,
 & &
 p_{n,s} = - \frac{1}{2} y\left(y-1\right)^{1+s}\left(y-9\right) \left( \frac{\psi^{(a)}_1}{\pi} \right)^n.
\eq
As an example consider the second graph in fig.~\ref{fig456}.
We have
\bq
 J_{14}
 & = &
 \eps^3 \left(1-y\right) I_{1102001}
 \nonumber \\
 & = &
 \left[ 
 - I\left(p_{3,0},1,p_{3,0};\tau_6^{(a)}\right)
 - 2\zeta_2 I\left(p_{3,0};\tau_6^{(a)}\right)
 \right] \eps^3
 + \left[
 I\left(p_{3,0},1,f_2,p_{3,0};\tau_6^{(a)}\right)
 \right. \nonumber \\
 & & \left.
 + I\left(p_{3,0},f_2,1,p_{3,0};\tau_6^{(a)}\right)
 + 2 \zeta_2 I\left(p_{3,0},f_2;\tau_6^{(a)}\right)
 - 2 \zeta_2 I\left(p_{3,0},1;\tau_6^{(a)}\right)
 \right. \nonumber \\
 & & \left.
 - \left( 7 \zeta_3 - 12 \zeta_2 \ln\left(2\right) \right) I\left(p_{3,0};\tau_6^{(a)}\right)
 \right] \eps^4
 + O\left(\eps^5\right),
\eq
with $f_2= - g_{2,0}/2 + g_{2,1} + g_{2,9}$
and $\tau_6^{(a)} = \psi_2^{(a)}/(6 \psi_1^{(a)})$, with
$\psi_1^{(a)}$ and $\psi_2^{(a)}$ being the periods of the elliptic curve associated to the sunrise graph.

Finally, there are integrals, which depend on $s$ and $t$. 
\begin{figure}
\begin{center}
\includegraphics[scale=1.0]{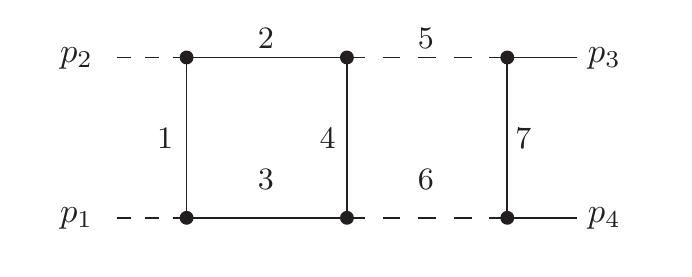}
\includegraphics[scale=1.0]{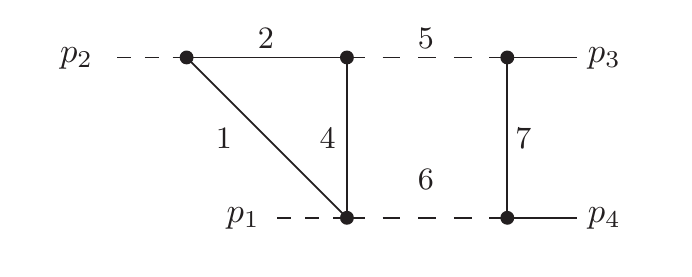}
\includegraphics[scale=0.8]{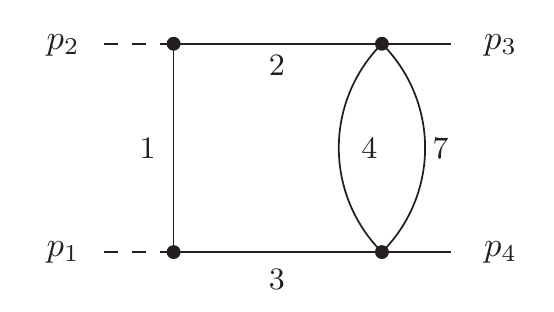}
\includegraphics[scale=0.8]{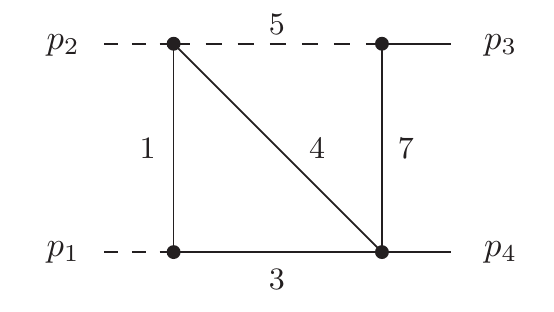}
\includegraphics[scale=0.8]{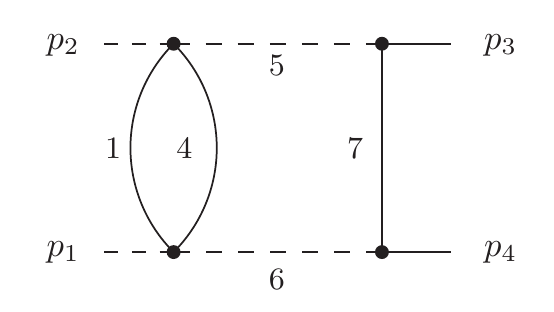}
\end{center}
\caption{
Feynman integrals depending on $s$ and $t$.
}
\label{fig7}
\end{figure}
These are shown in fig.~\ref{fig7}.
In order to construct the basis $\vec{J}$ for these topologies we first consider the diagonal blocks.
For the diagonal blocks we combine the information
from the maximal cuts with the technique based on the factorisation properties 
of Picard-Fuchs operators \cite{Adams:2017tga}.
For the non-diagonal blocks we use a modified version of the algorithm of Meyer \cite{Meyer:2016slj,Meyer:2017joq}.
Let us first give an example on how to exploit the factorisation properties of the Picard-Fuchs operator.
In sector 123 (the upper right graph in fig.~\ref{fig7}) we have two master integrals, however the Picard-Fuchs operator
factorises in even integer dimensions. From the factorisation we construct two master integrals
\bq
 J_{38}
 & = & 
 2 \eps^4 
 {\frac { \left( {x}^{2}+1 \right)  \left( {x}^{2}-2\,x-1 \right) }{ \left( x-1 \right)  \left( x+1 \right) x}}
 \left[ I_{11011110(-1)} - \left(y-2\right) I_{1101111} \right]
 - 4\,{\frac {x+1}{{x}^{2}+1}} J_{22},
 \nonumber \\
 J_{39}
 & = & 
 \eps^4 \left(1-y\right) {\frac { \left( {x}^{2}+1 \right) ^{2}}{ \left( x-1 \right)  \left( x+1 \right) x}} I_{1101111},
\eq
which bring the diagonal block into a $\eps$-form.
$I_{11011110(-1)}$ denotes a Feynman integral with a numerator.
In all other topologies of fig.~\ref{fig7} we find a factorisation involving exactly one second-order irreducible factor supplemented by
additional first-order factors.
In particular, no irreducible differential operator of order three or higher occurs.
In the next step we identify the associated elliptic curves.
We recall that in the sunrise integral an elliptic curve can either be obtained from the Feynman graph polynomial
or the maximal cut \cite{Adams:2013nia,Adams:2014vja}.
The periods $\psi_1^{(a)}$ and $\psi_2^{(a)}$ of the elliptic curve are solutions of the homogeneous differential equation.
It is further known, that the maximal cuts are always solutions of the homogeneous differential equations \cite{Primo:2016ebd}.
We therefore search for Feynman integrals, whose maximal cuts are periods of an elliptic curve.
The analysis of the maximal cuts is most easily carried out in the Baikov representation \cite{Baikov:1996iu,Lee:2009dh,Kosower:2011ty,CaronHuot:2012ab,Frellesvig:2017aai,Bosma:2017ens,Harley:2017qut}.
As an example we consider the maximal cuts of the sunrise integral and the double box integral:
\bq
\lefteqn{
 \mathrm{MaxCut}_{\mathcal C} \; I_{1001001}\left(2-2\eps\right)
 = } & & \\
 & &
 \frac{u m^2}{\pi^2}
 \int\limits_{\mathcal C} 
 \frac{dP}{\left(P -t \right)^{\frac{1}{2}} \left(P - t + 4 m^2 \right)^{\frac{1}{2}} \left(P^2 + 2 m^2 P - 4 m^2 t + m^4\right)^{\frac{1}{2}}}
 +
 {\mathcal O}\left(\eps\right),
 \nonumber \\
\lefteqn{
 \mathrm{MaxCut}_{\mathcal C} \; I_{1111111}\left(4-2\eps\right)
 = } & & \nonumber \\
 & &
 \frac{u m^6}{4 \pi^4 s^2}
 \int\limits_{\mathcal C} 
 \frac{dP}{\left(P -t \right)^{\frac{1}{2}} \left(P - t + 4 m^2 \right)^{\frac{1}{2}} \left(P^2 + 2 m^2 P - 4 m^2 t + m^4 - \frac{4m^2\left(m^2-t\right)^2}{s} \right)^{\frac{1}{2}}}
 +
 {\mathcal O}\left(\eps\right),
 \nonumber
\eq
where $u$ is an (irrelevant) phase and the contour ${\mathcal C}$ is between two points, where the denominator vanishes.
From the denominator we may now easily read off the elliptic curve.
Repeating this for all elliptic sectors, we find three different elliptic curves:
\bq
 E^{(a)} & : & w^2 = \left(z -t \right) \left(z - t + 4 m^2 \right) \left(z^2 + 2 m^2 z - 4 m^2 t + m^4\right),
 \\
 E^{(b)} & : & w^2 = \left(z -t \right) \left(z - t + 4 m^2 \right) \left(z^2 + 2 m^2 z - 4 m^2 t + m^4 - \frac{4m^2\left(m^2-t\right)^2}{s} \right),
 \nonumber \\
 E^{(c)} & : & w^2 = \left(z -t \right) \left(z - t + 4 m^2 \right) \left(z^2 + \frac{2 m^2 \left(s+4t\right)}{\left(s-4m^2\right)} z
           + \frac{s m^2 \left(m^2-4t\right) - 4 m^2 t^2}{s-4m^2} \right).
 \nonumber
\eq
The curve $E^{(a)}$ is associated to the sunrise integral, the curve $E^{(b)}$ is associated to the double box integral and sectors $79$ and $93$
(the bottom left and bottom middle graphs in fig.~\ref{fig7}),
the curve $E^{(c)}$ is associated to sector $121$ (the bottom right graph in fig.~\ref{fig7}).
The curve $E^{(a)}$ gives rise to iterated integrals of modular forms of $\Gamma_1(6)$.
It is easy to see that the curves $E^{(b)}$ and $E^{(c)}$ degenerate to $E^{(a)}$ for $s\rightarrow \infty$.
However, for $s \neq \infty$ they are distinct.
If we would have only one curve, we expect that the result can be written in elliptic polylogarithms \cite{Brown:2011,Broedel:2017kkb},
which are iterated integrals on a single elliptic curve.
Let us stress that we have three elliptic curves.

We continue with the construction of the master integrals.
From the sunrise sector it is known, that we may choose one master integral as the one having the right maximal cut, normalised by its
maximal cut.
The second master integral related to the irreducible second-order differential operator can then be chosen as a linear
combination of this integral and its derivative (and sub-topologies).
If the topology has more than two master integrals, these two master integrals are supplemented by additional master integrals related
to the first-order differential operators.
This pattern applies to all elliptic sectors.
As an example we consider the sector $79$. We have three master integrals, which can be chosen as
\bq
 J_{24}
 & = \;\; & 
 \eps^3 
 \frac{\left(1+x^2\right)^2}{x\left(1-x^2\right)}
 \frac{\pi}{\psi^{(b)}_1} I_{1112001},
 \\
 J_{25}
 & = \;\; & 
 \eps^3 \left(1-2\eps\right) 
 \frac{\left(1+x^2\right)^2}{x\left(1-x^2\right)}
 I_{1111001}
 + R_{25,24} \frac{\psi^{(b)}_1}{\pi} J_{24},
 \nonumber \\
 J_{26}
 & = \;\; & 
 \frac{6}{\eps} \frac{\left(\psi^{(b)}_1\right)^2}{2 \pi i W^{(b)}_y} \frac{d}{dy} J_{24}
 + R_{26,24} \left( \frac{\psi^{(b)}_1}{\pi} \right)^2 J_{24}
 - \frac{\eps^2}{24} \left(y^2-30y-27\right) \frac{\psi^{(b)}_1}{\pi} {\bf D}^- I_{1001001},
 \nonumber
\eq
where $\psi^{(b)}_1$ denotes a period of the curve $E^{(b)}$, $W^{(b)}_y$ the Wronskian and $R_{25,24}$ and $R_{26,24}$ rational functions in $(x,y)$.
We thus arrive at the differential equation
\bq
 d \vec{J}
 & = &
 \left( A^{(0)} + \eps A^{(1)} \right) \vec{J},
\eq
where $A^{(0)}$ is strictly lower triangular and
$A^{(1)}$ is as usual block triangular.
Furthermore, $A^{(0)}$ vanishes for $t=m^2$ or $s=\infty$.
In addition, $A^{(1)}$ reduces to one-forms associated with polylogarithms for $t=m^2$ and to modular forms for $s=\infty$.
The entries of $A^{(0)}$ and $A^{(1)}$ are rational in
\bq
\label{linear_in_eps_2}
 \left\{ \; x, y, \; \psi_1^{(a)}, \psi_1^{(b)}, \psi_1^{(c)}, \; \partial_y \psi_1^{(a)}, \partial_y \psi_1^{(b)}, \partial_y \psi_1^{(c)} \; \right\}.
\eq
The system of differential equations in eq.~(\ref{linear_in_eps_2}) is easily solved.
The full result is given in an auxiliary electronic file accompanying ref.~\cite{Adams:2018kez}.

\section{Conclusions}

Loop integrals with internal masses are important for top-, $W$/$Z$- and Higgs-physics at the LHC.
They may involve elliptic sectors from two loops onwards.
We showed in this talk that in the calculation of master integrals more than one elliptic curve may occur.
This is the case for the planar double box integral relevant to top-pair production with a closed top loop.
We also showed that despite this complication, the system of differential equations may be brought into a form linear in $\eps$, 
where the $\eps^0$-term is strictly lower triangular.
This system of differential equations is easily solved in terms of iterated integrals to any order in $\eps$.
We expect the methods discussed here to be useful for a wider class of Feynman integrals.

% -----------------------------------------------------------------------------
% references
\bibliography{/home/stefanw/notes/biblio}
\bibliographystyle{/home/stefanw/latex-style/h-physrev5}

\end{document}